\documentclass[preprint,showpacs,preprintnumbers,amsmath,amssymb]{revtex4}

\usepackage{graphicx}
\usepackage{dcolumn}
\usepackage{bm}
\usepackage{textcomp}

\begin{document}

\preprint{APS/123-QED}

\title{Detection mechanism for ferroelectric domain boundaries \\ with lateral force microscopy}

\author{Tobias Jungk}
\email{jungk@physik.uni-bonn.de}
\author{\'{A}kos Hoffmann}
\author{Elisabeth Soergel}

\affiliation{Institute of Physics, University of Bonn,
Wegelerstra\ss e 8, 53115 Bonn, Germany}

\date{\today}

\begin{abstract}
The contrast mechanism for the visualization of ferroelectric domain
boundaries with lateral force microscopy is generally assumed to be
caused by mechanical deformation of the sample due to the converse
piezoelectric effect. We show, however, that electrostatic
interactions between the charged tip and the electric fields arising
from the surface polarization charges dominate the contrast
mechanism. This explanation is sustained by quantitative analysis of
the measured forces as well as by comparative measurements on
different materials.
\end{abstract}

\pacs{77.80.Dj, 68.37.Ps, 77.84.-s, 84.37.+q}

\maketitle


Lateral force microscopy (LFM) is usually utilized for the detection
of friction forces between tip and sample while scanning the
surface. LFM can therefore map topographical steps (roughness) but
also different material compositions, whereby the contrast depends
strongly on the scanning parameters (velocity, direction etc.)
\cite{Wie}.

In first LFM experiments with ferroelectric samples, investigating
GASH and TGS, such dependences were observed, indicating that the
domain contrast was due to topographical features, as these
materials exhibit domain selective, hygroscopic properties
\cite{Lue93,Gru95}. Domains or domain boundaries have also been
revealed by LFM in other crystals such as $\rm LiNbO_3$
\cite{Wit02,Scr05}, $\rm KTiOPO_4$ \cite{Wit02,Can03}, $\rm BaTiO_3$
\cite{Eng98}, and PZT \cite{Rod04,Rod05}, using the very same
experimental setup as for piezoresponse force microscopy (PFM)
\cite{Alexe}. In brief, an alternating voltage (amplitude $V_{\rm
AC}$; frequency $f_{\rm AC}$) is applied to the tip which leads to
deformations of the sample due to the converse piezoelectric effect.
The resulting vibrations of the surface cause oscillations of the
cantilever which can be read out with a lock-in amplifier. For LFM,
instead of the bending, the torsion of the cantilever is detected.
In the case of anti-parallel $c$-domains,
the alternating voltage leads to tilting vibrations of the surface
at the domain boundaries. This tilting is supposed to cause lateral
forces acting on the tip which result in torsional oscillations of
the cantilever \cite{Wit02,Scr05}. Our investigations show, however,
that these torsional oscillations are dominated by the electrostatic
interaction between the charged tip and the electric fields arising
from the surface polarization charges. We therefore name this
detection technique lateral electrostatic force microscopy (LEFM).

For the experiments we used a commercial scanning force microscope
(Smena, NT-MDT), modified to allow application of voltages to the
tip. The AC-voltage driven bending and torsion of the cantilever
were recorded simultaneously as vertical and lateral response,
respectively. To obtain accurate experimental data, the alignment of
the four-segmented photo-detector with respect to the cantilever is
crucial. In case of a misalignment, a pure bending of the cantilever
inevitably pretends a (nonexisting) torsion, and vice versa. The
magnitude of this cross-talk can be estimated when exciting the
cantilever in air at its resonance frequency and comparing the
vertical and the lateral response: We measured cross-talks of
typically 5 - 20\% why we developed a cross-talk compensator which
electronically suppresses the cross-talk by a factor of 100.
Therefor we add a 180$^{\circ}$ phase-shifted correction-signal of
adjustable amplitude to the error-signal. Note that for every
cantilever, the compensation has to be adjusted anew.

\begin{figure}
\includegraphics{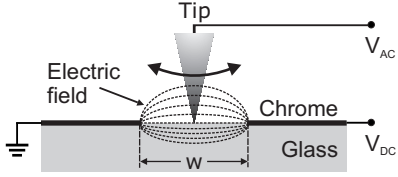}
\caption{\label{fig:Jungk1} Setup for the LEFM calibration: a chrome
mask with a slit of $w = 20$\,\textmu m width allows to generate an
electric field parallel the surface by applying a voltage $V_{\rm
DC}$ to the electrodes. Simultaneously, an alternating voltage
$V_{\rm AC}$ is applied to the tip. The amplitude and phase of the
enforced oscillations of the tip can be read out with a lock-in
amplifier.}
\end{figure}

The LEFM setup was calibrated with a sample consisting of a glass
plate with chrome electrodes which allow to generate electric fields
parallel to the sample surface (Fig.~\ref{fig:Jungk1}). The chrome
mask had a thickness of 250\,nm and a slit width of $w=20$\,\textmu
m. Applying a voltage of $V_{\rm DC} = 20$\,V to the electrodes
generates an electric field of $E\approx V_{\rm DC}/w = 10^6$\,V/m
within the slit which at its center is mainly parallel to the
surface. The tip ($V_{\rm AC} = 10$\,V$_{\rm pp}$; $f_{\rm AC}\sim
33$\,kHz) senses this electric field $E$ and performs oscillations
which can be read out with a lock-in amplifier. The positioning of
the tip was found to be uncritical along a section of at least
10\,\textmu m width in the middle of the slit.
This setup allows to determine the phase of the oscillations with
respect to the direction of the electric field. Furthermore, the
influence of friction between tip and sample surface on the
oscillation amplitude can be measured: Altering the load up to
30\,\textmu N  had no influence on the signal, however, when
retracting the tip, an enlargement by a factor of 1.5 was measured.

In LFM the orientation of the cantilever with respect to the lateral
force $F_{\rm l}$ acting on the tip is crucial. If $F_{\rm l}$ is
perpendicular to the cantilever, the latter will be forced to twist
which can be read out as a torsion signal. If, on the contrary,
$F_{\rm l}$ is orientated parallel to the cantilever, this will lead
to a buckling, which can be detected as a deflection signal. The LFM
signals were calibrated taking into account the appropriate force
constants of the cantilever and assuming that the photo detector has
the same sensitivity for vertical as for lateral signals. Be aware
that for topographical images the contrast information is height [m]
whereas for LFM images it is force [N].
The dimensions of the cantilevers used for the measurements are:
length $l$~=~130\,\textmu m, width $w$~=~35\,\textmu m, thickness
$t$~=~1.9\,\textmu m, and tip height $h$~=~17\,\textmu m. Therefore
the lateral spring constants are $k_{\rm t}= (Gwt^3)/(3lh^2) \approx
95.5$\,N/m for torsion and $k_{\rm b}= (Ewt^3)/(12lh^2) \approx
90$\,N/m for buckling (shear modulus $G=0.5\times 10^{11}$\,N/m$^2$
and Young's elasticity modulus $E=1.7\times 10^{11}$\,N/m$^2$
\cite{Lide}). For the sake of completeness we give the deflection
spring constant which is $k_{\rm d} = (Ewt^3)/(4l^3) \approx
4.6$\,N/m.

\begin{figure}
\includegraphics{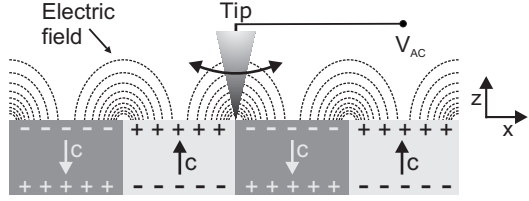}
\caption{\label{fig:Jungk2} Schematic drawing of the static electric
fields above the $z$ face of a periodically poled ferroelectric
crystal. Here $c$ denotes the optical axis, $V_{\rm AC}$ the
alternating voltage applied to the tip.}
\end{figure}

We investigated a $z$-cut periodically-poled $\rm LiNbO_3$ crystal
(PPLN) with a period length of 8\,\textmu m and a thickness of
0.5\,mm. The appropriate piezoelectric coefficient is known to be
$d_{33}=7.6$\,pm/V and the (uncompensated) surface polarization
charge density is $\sigma = 0.71$\,C/m$^2$ \cite{Landolt}.
Figure~\ref{fig:Jungk2} shows a sidewise sketch of a PPLN crystal.
Because of the surface polarization charges, electric fields build
up whose strength parallel to the surface is most at the domain
boundaries. The electric field $E_x(x,z)$ with  $x$ being the axis
parallel to the surface and perpendicular to the domain boundaries,
and $z$ denoting the distance from the sample surface
(Fig.~\ref{fig:Jungk2}) for an infinite PPLN structure is given by
\begin{equation*}
E_x(x,z) = \frac{\sigma}{4\pi\varepsilon_0}\,\ln\left[\; \prod_{n =
-\infty}^{\infty}\; \frac{\big[(x + 2na)^2 + z^2\big]^2}{\big[(x +
2na + a)^2 + z^2 \big]^2}  \; \right]
\end{equation*}
with $a$ denoting the domain size (PPLN period: $2a$) and $n$ the
number of domains being included. For the PPLN sample electric field
strengths of $10^{11}$\,V/m are theoretically expected if no
compensation of the surface charges is assumed.

\begin{figure}
\includegraphics{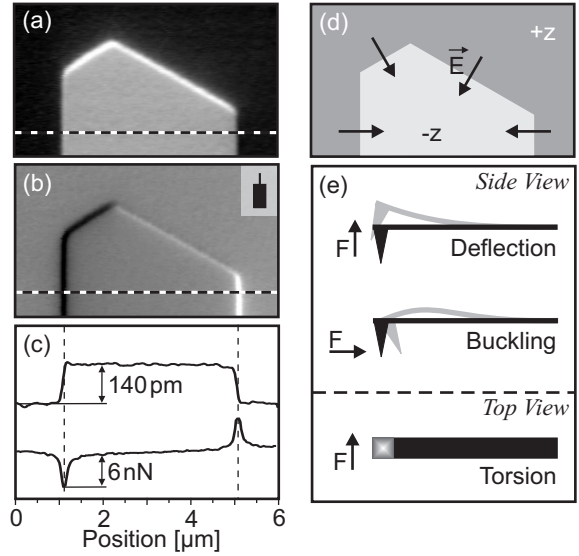}
\caption{\label{fig:Jungk3} Deflection (a) and torsion (b) images
simultaneously recorded on a $\rm LiNbO_3$ crystal (image size $6
\times 3.5$\,\textmu m$^2$) with the corresponding scanlines (c)
with 10\,V$_{\rm pp}$ applied to the tip. The orientation of the
chip with the cantilever is shown as inset in (b).
Schematic drawing (d) of the electric field distribution $\vec E$.
In (e) the possible movements of the cantilever are depicted. The
deflection image (a) shows deflection (PFM) and buckling (LEFM), the
torsion image (b) the twisting of the cantilever (LEFM).}
\end{figure}

In Fig.~\ref{fig:Jungk3} the experimental results for deflection (a)
and torsion (b) images of the end of a poled stripe of PPLN are
shown, with the corresponding scan lines in (c). The orientation of
the cantilever was chosen to be parallel to the stripe (see also
inset of Fig.~\ref{fig:Jungk3}(b)). At first sight it is obvious
that the deflection image (a) shows the domain faces (due to the
converse piezoelectric effect) whereas the torsion image (b) only
shows the domain boundaries, at the left edge as a dark stripe and
at the right edge as a bright stripe. The contrast inversion is due
to the change of the direction of the electric field  (see also
Fig.~\ref{fig:Jungk3}(d)). This is consistent with the results we
obtained with the test sample from Fig.~\ref{fig:Jungk1}, where we
determined the sign of the output of the lock-in amplifier with
respect to the direction of the electric field. The contrast is
reduced when the electric field vector perpendicular to the
cantilever becomes smaller as it can be seen on the tilted edges of
the domain. As the cross-talk between vertical and lateral signals
was suppressed, the level of the torsion signal within and outside
the domain is same.
Looking more closely at Fig.~\ref{fig:Jungk3}(a), at the top edges
of the domain a bright stripe is visible. When comparing with the
schematic drawing of the electric field configuration in
Fig.~\ref{fig:Jungk3}(d) at these edges the electric field has a
component along the axis of the cantilever. This also leads to
lateral forces acting on the tip which result in a buckling of the
cantilever.

To summarize the experimental results of Fig.~\ref{fig:Jungk3}: the
deflection image Fig.~\ref{fig:Jungk3}(a) shows the vertical
movement of the cantilever due to the converse piezoelectric effect
and, at the top edges the electrostatic interaction of the
periodically charged tip with the electric field which leads to a
buckling of the cantilever (upper part of Fig.~\ref{fig:Jungk3}(e)).
The torsion image Fig.~\ref{fig:Jungk3}(b) only shows the electric
field component perpendicular to the axis of the cantilever (lower
part of Fig.~\ref{fig:Jungk3}(e)).

To support our explanation, that lateral forces measured at the
domain boundaries are due to electrostatic interactions of the
periodically charged tip with the electric field arising from the
surface polarization charges, we analyzed the data quantitatively.
Assuming a tip radius of $r=30$\,nm and a voltage of $U=
10$\,V$_{\rm pp}$ applied to it results in a charge of $Q = U\,
4\pi\varepsilon r \approx 10^{-17}$\,C. In an external electric
field $E=10^{11}$\,V/m this leads to a force $F=Q \, E \approx
10^{-6}$\,N. As the strength of the electric field $E$ is directly
proportional to the surface charge density $\sigma$, comparing this
result with the measured value of $F=6$\,nN 
underlines that the surface charge is reduced by three orders of
magnitude by compensation charge which agrees with other
publications \cite{Kal01,Lik01}.

For comparison, we estimated the lateral forces expected from the
tilting of the surface due to the converse piezoelectric effect.
From the tip size and the lateral resolution in PFM measurements one
can assume that the tilt of the surface occurs in a region of
$\Delta x = 100$\,nm diameter. The maximum vertical surface
displacement in $\rm LiNbO_3$ is $\Delta h = 2 \times 76$\,pm with
$V_{\rm AC}= 10\,V_{\rm pp}$ applied to the tip. The surface can
then be described as a inclined plane with an angle of $\alpha
 = 0.9 \times 10^{-3}[^{\circ}]$. A vertical force $F_{\rm v}$
acting on this inclined plane can be vectorially devised in two
components: one normal to the slanted surface ($F_{\rm n}$) and the
other one parallel to it ($F_{\rm l}$) causing a torsion of the
cantilever. From simple geometrical considerations it is evident
that $\Delta h/\Delta x = F_{\rm l}/F_{\rm n} \approx F_{\rm
l}/F_{\rm v}$ because of $\alpha$ being very small. When operating a
scanning force microscope in contact mode, the typical load of the
tip acting on the surface is $F_{\rm v}\approx 10$\,nN \cite{Wie}
and therefore the expected lateral forces are $F_{\rm l}<
0.015$\,nN. This is smaller by almost three orders of magnitude than
the values measured e.g. in Fig.~\ref{fig:Jungk3}(b). Therefore the
mechanical contribution to the lateral forces is negligible.
Furthermore, we have not observed any dependence of the lateral
forces measured on the load of the tip, as it is required for this
contrast mechanism.

Finally we carried out comparative LEFM measurements on $\rm
LiNbO_3$ and SBN crystals. Because the piezoelectric coefficient of
SBN is three times larger than of $\rm LiNbO_3$, the expected
tilting of the surface at the domain boundaries should be much
steeper. The measured lateral forces, however, are smaller by a
factor of 6 with respect to those on $\rm LiNbO_3$. This agrees well
with an electrostatic origin of the lateral forces as the surface
polarization charge density is smaller for SBN than for $\rm
LiNbO_3$.

In conclusion, we have shown that the origin of the contrast
mechanism for the detection of domain boundaries in lateral force
microscopy is dominated by the electrostatic interaction of the
charged tip with the electric field arising from the surface
polarization charges. A quantitative estimate as well as comparative
measurements on $\rm LiNbO_3$ and SBN crystals sustain this
explanation.


\begin{acknowledgments}
We thank Boris Sturman for fruitful discussions. Financial support
of the DFG research unit 557 and of the Deutsche Telekom AG is
gratefully acknowleged.
\end{acknowledgments}


\end{document}